\documentstyle[preprint,aps]{revtex}
\begin {document}
\draft
\preprint{UCI TR 93-28}
\title{Fields in the Vicinity of a Superconducting Cosmic String}
\setcounter{footnote}{0}
\author{Myron Bander\footnote{Electronic address:
mbander@funth.ps.uci.edu}\addtocounter{footnote}{3}%
}
\address{
Department of Physics, University of California, Irvine, California
92717, USA}
\author{H. R. Rubinstein\footnote{Electronic address: rub@vand.physto.se}}
\address{
Department of Radiation Sciences, University of Uppsala, Uppsala,
Sweden}
\date{August\ \ \ 1993}
\maketitle
\begin{abstract}
Superconducting cosmic strings may be viewed as wires
of thickness $1/\Lambda$ with $\Lambda=10^{16}$ TeV. We show that the weak
interactions will spread out the current to distances $r= (1/M_Z)\ln
(I/M_Z)$, where $I$ is the magnitude of the current in the string.
Consequences for the scattering of light by these strings is presented.
\end{abstract}

\narrowtext
Symmetry breaking at scales of
$\Lambda=10^{16}$ TeV, or higher, may induce strings that behave as
superconducting wires \cite{Witten} carrying currents of the order of
$I=10^{20}$ A. The radius of these ``wires'' is governed by the masses of
the Higgs particles responsible for the symmetry breaking and will be of
the order of $1/\Lambda$. It has been noted that, due to other
interactions, various instabilities will develop at  larger radii.
Hadronic chiral symmetry breaking will screen fields for $r\le I/f_\pi
m_\pi$ \cite{BanRub}; $r$ is the distance from the wire. At distances of
the order of $r=I/M_W^2$ the anomalous magnetic moment of the $W$ boson
induces a condensation \cite{Ambjorn} of the fields corresponding to this
particle. Although the screening discussed in Ref.~\cite{BanRub} extends
to the largest distances considered so far, the mechanism is suspect in
that it relies on a low energy effective model rather than a fundamental
theory. In this work we shall show that at even smaller distances, $r\le
(1/M_Z)\ln (I/M_Z)$, the magnetic field is partially screened by the weak
interactions. The definition of an electric current in the presence of a
non Abelian gauge theory with the electromagnetic field being one of the
adjoint fields is problematic as a local gauge transformation can rotate
the current to an other direction in group space. We shall return to this
point shortly.

As in the $SU(2)\times U(1)$ theory of the weak interactions the
electromagnetic field is a combination of an $SU(2)$ and a $U(1)$
field we shall, for pedagogical reasons, start with the Georgi-Glasgow
$O(3)$ model \cite{GG} and then return to the full Weinberg-Salam theory.
The fields of the $O(3)$ model are $W_\mu^{(1)}$,  $W_\mu^{(2)}$ and
$A_\mu =W_\mu^{(3)}$. In addition there is a triplet of Higgs fields
$\phi^{(i)},\ i=1,2,3$. Again, for simplicity we shall work in the
nonlinear limit where $\phi^2$ is a constant and scaled to equal one.  The
Lagrangian for these fields coupled to an external electric current
$j^\mu$ is  \begin{equation}
{\cal L}=-\frac{1}{4}F^{(i)}_{\mu\nu}F^{(i)\mu\nu}+ \frac{v^2}{2}
(\partial_\mu\phi^{(i)}+g\epsilon^{ijk}W_\mu^{(j)}\phi^{(k)})
(\partial^\mu\phi^{(i)}+g\epsilon^{ijk}W^{\mu (j)}\phi^{(k)})
+j^\mu A_\mu\, ;\label{lag1}
\end{equation}
$g$ is the coupling constant and $v$ is the vacuum expectation value of
the Higgs field. The current $j$ will be taken as that due to a wire at
the origin and extending along the z direction, $j_z=gI\delta
(\mbox{\boldmath $r$})$, with \mbox{\boldmath $r$} the spatial vector
transverse to the current direction. We may perform a gauge rotation, so
that the Higgs field points everywhere in the $3$ direction,
\begin{eqnarray}
{\cal L}=&-&\frac{1}{4}F^{(i)}_{\mu\nu}F^{(i)\mu\nu}+\frac{M^2}{2}
\left(W_\mu^{(1)}W^{\mu(1)}+W_\mu^{(2)}W^{\mu(2)}\right)\nonumber\\
&-&gI\delta (\mbox{\boldmath $r$})
\left[O^{3i}W_z^{(i)}+\frac{1}{g}\epsilon
^{ijk}\partial_zO^{3j}O^{3k}\right]\, ;\label{lag2}
\end{eqnarray}
$M=gv$ is the mass of the charged vector mesons.

As we shall be interested in the static energies of configurations that
are solutions of the equations of motion derived from Eq.~(\ref{lag2}),
it is the Hamiltonian we need;
\begin{eqnarray}
{\cal H}&=&\int d^2r\frac{1}{4}\left [\partial_a A_b-\partial_b
A_a+g(W^{(1)}_aW^{(2)}_b- W^{(1)}_bW^{(2)}_a)\right ]^2\nonumber\\
&+&\frac{1}{4}\left[\partial_a W^{(i)}_b-\partial_b
W^{(i)}_a+g\epsilon^{ij}(W^{(j)}_a A_b-W^{(j)}_bA_a)\right]^2\nonumber\\
&+&gI\delta (\mbox{\boldmath $r$})
\left[O^{3i}W_z^{(i)}+\frac{1}{g}\epsilon
^{ijk}\partial_zO^{3j}O^{3k}\right]\, ;\label{ham1}
\end{eqnarray}
$a,\ b$ are the spatial directions and $\epsilon^{ij}$ is the
two dimensional Levi-Civita symbol. We wish to minimize the
energy with respect to the variables $A_a,\ W^{(i)}_a,\ i=1,2$ and
$O^{ij}$. An obvious candidate is the solution corresponding to
classical electromagnetism,
\begin{equation}
A_z=\frac{gI}{2\pi}\ln\frac{r}{r_0}\, ,\label{classA}
\end{equation}and all other fields set equal to zero; $r_0$ is a gauge
parameter which may be set equal to the radius of the wire. The energy
per unit length along the wire that is contained inside a cylinder of
radius $R$ is
\begin{equation}
{\cal E}=\frac{g^2I^2}{4\pi}\ln\frac{R}{r_0}\, .\label{classE}
\end{equation}
A configuration whose energy does not have the logarithmic dependence on
$R$ can be obtained by rotating the current into one of the massive
directions, say we let it couple to $W^{(1)}_z$; in this case only
this field is excited and we find
\begin{equation}
W^{(1)}_z=-\frac{I}{2\pi}K_0(Mr)\, ,\label{naivconf}
\end{equation}
and the corresponding energy per unit length is
\begin{equation}
{\cal E}=-\frac{g^2I^2}{4\pi}K_0(Mr_0)\, ,
\end{equation}
which is certainly lower than that of Eq.~(\ref{classE}). This solution
is however unacceptable as it is not what we would interpret as the
fields resulting from an electric current.

We take as the definition of a configuration resulting from an
electric current along the z-axis to be one where the magnetic field
satisfies Amp\`ere's law at large distances from the wire
\begin{equation}
\oint\mbox{\boldmath $B\cdot dl$}=\frac{gI}{2\pi}\, ,\label{Ampere}
\end{equation}
with the integral taken around a contour far from the wire. Certainly,
the configuration described by Eq.~(\ref{classA}) satisfies this
criterion while that of Eq.~(\ref{naivconf}) {\em does not\/}. There
is however a configuration that does satisfy Eq.~(\ref{Ampere}) and
has a lower energy than that of  Eq.~(\ref{classE}).

As for the case of Eq.~(\ref{naivconf}) we take the rotation to be
such as to couple the current to $W^{(1)}_z$ and look for
configurations satisfying Eq.~(\ref{Ampere}). We also find that only
$A_z,\ W^{(1)}_z$ and $W^{(2)}_{x,y}$ are excited. In order to
simplify notation, we introduce
\begin{eqnarray}
A_z&=&A\, ,\nonumber\\
W^{(1)}_z&=&W\, ,\\
W^{(2)}_{x,y}&=&V_{x,y}\, .\nonumber
\end{eqnarray}
In terms of these variables the Hamiltonian is
\begin{eqnarray}
{\cal H}&=&\int d^2r\left [\frac{1}{2}(\partial_aA+gWV_a)^2+
\frac{1}{2}(\partial_aW-gAV_a)^2\right. \nonumber\\
 &+&\left. \frac{1}{2}(\partial_aV_b-\partial_bV_a)^2+\frac{M^2}{2}(W^2+V_a^2)
+gWI\delta (\mbox{\boldmath $r$})\right ]\, .\label{ham2}
\end{eqnarray}
As all these fields will depend only on the radial distance $r$ and
the field $V_a$ will point in the radial direction,
the term involving the derivatives of this field will vanish.
Defining new scaled variables
\begin{equation}
\rho=g^2Ir\, ,\ \ \mu=\frac{M}{g^2I}\, ,\ \ a=\frac{A}{gI}\, ,\ \
w=\frac{W}{gI}\, ,\ \  v=\frac{V_r}{gI}\, .
\end{equation}
The energy is
\begin{equation}
{\cal H}=g^2I^2\int d^2\rho\left [
\frac{1}{2}(\partial_\rho a+wv)^2+\frac{1}{2}(\partial_\rho w-av)^2+
\frac{\mu^2}{2}(w^2+v^2)+w\delta(\mbox{\boldmath $\rho$})\right ] \, .
\end{equation}
As no kinetic energy terms appear for $v$ its equation of motion may
be solved
\begin{equation}
v=\frac{w\partial_\rho a-a\partial_\rho w}{\mu^2+w^2+a^2}
\end{equation}
resulting in
\begin{equation}
{\cal H}=g^2I^2\int d^2\rho\Bigg [\frac{1}{2}(\partial_\rho a)^2+
\frac{1}{2}(\partial_\rho w)^2+\frac{\mu^2}{2}w^2
-\frac{1}{2}\frac{(w\partial_\rho a
-a\partial_\rho w)^2}{\mu^2+w^2+a^2}+
w\delta(\mbox{\boldmath $\rho$})\Bigg ]\, .\label{ham3}
\end{equation}
The equations obtained from this Hamiltonian are far too non-linear in
order to be able to obtain analytic solutions. We shall, instead,
use the variational principle and show a solution with the property
demanded by Eq.~(\ref{Ampere}) and whose energy is lower than that of
Eq.~(\ref{classE}).
\begin{eqnarray}
a&=&\left\{\begin{array}{ll}
           0   & \mbox{for $\rho<\rho_1$}\\
           \frac{I}{2\pi}\ln (\rho /\rho_1)\ \  &  \mbox{for
$\rho>\rho_1$}
          \end{array}
         \right. \nonumber\\
&{}&\label{varsol}\\
w&=&-\frac{1}{2\pi}K_0(\mu\rho)\, ;\nonumber
\end{eqnarray}
$\rho_1$ is a parameter to be determined by minimizing the energy. The
difference in the energies $\delta ({\cal E})$ of the purely
electromagnetic case, Eq.~(\ref{classE}) and the one due to the above
configuration is (for  $\mu\rho_1 > 1$)
\begin{equation}
\delta ({\cal E})=\frac{1}{4\pi}\ln (\mu\rho_1)+\pi\int_{\rho_1} \rho
d\rho \frac{(w\partial_\rho a-a\partial_\rho w)^2}{\mu^2+w^2+a^2}\, .
\label{delta}
\end{equation}
$\delta ({\cal E})$ is positive for all $\rho_1$'s and thus we have found
configurations satisfying Amp\`ere's law and whose energies are lower
than that of the purely electromagnetic case.

It remains to find the optimal value of $\rho_1$. We are interested in
large currents, therefore, for fixed vector meson mass $M$ this means
small $\mu$. The full minimization of Eq.~(\ref{delta}) cannot be
done in an analytic form but an asymptotic one,
valid for small $\mu$, can be obtained
\begin{equation}
\mu\rho_1=-\ln(\mu)-\frac{1}{2}\ln\left [\frac{-2\ln(\mu)}{\pi}\right ]
\, .\label{apprsol}
\end{equation}
A comparison with a numerical minimization of Eq.~(\ref{delta}) shows
that this solutions valid for $\mu\le 0.5$.  We thus find that the
magnetic field is screened for distances smaller than $r_1$ with
\begin{equation}
r_1=\frac{1}{M}F\left(\frac{g^2I}{M}\right )\, .
\end{equation}
For large currents $F=\ln(g^2I/M)$. This is the result presented at
the beginning of this work.

In the full Weinberg-Salam model the electromagnetic field is a
combination of the hypercharge field $B$ and the third component of the
weak isospin triplet $W^{(3)}$,
\begin{equation}
{\cal L}=\cdots +ej(\cos\theta_WB+\sin\theta_W W^{(3)})\, .
\end{equation}
It is only the isospin part that can be rotated by an $SU(2)$
transformation into a massive direction,
\begin{equation}
{\cal L}=\cdots +ej\left [\cos\theta_WB+\sin\theta_W
(\cos\frac{\beta}{2}W^{(3)}+\sin\frac{\beta}{2}W^{(1)})\right ]\, .
\end{equation}
Re expressed in terms of the mass eigenstates $A$ and $Z$ this becomes
\begin{eqnarray}
{\cal L}&=&\cdots +ej\left \{\sin\theta_W\sin\frac{\beta}{2}W^{(1)}
\left [(\cos\theta_W)^2+(\sin\theta_W)^2\cos\frac{\beta}{2}\right
]A\right. \nonumber\\
&+&\left. \cos\theta_W \sin\theta_W (1-\cos\frac{\beta}{2})Z\right \}\, .
\end{eqnarray}
The lowest energy configuration is obtained when the coefficient of the
electromagnetic field is as small as possible. This is achieved for
$\beta=2\pi$, which also eliminates the $W^{(1)}$ term. Thus out to
a distance
\begin{equation}
r_1=\frac{1}{M_Z}\ln\frac{I}{M_Z}
\end{equation}
$\cos 2\theta_W$ of the electromagnetic field will penetrate and the
$Z$ field will appear with a coupling to the current of $e\sin
2\theta_W$. Beyond $r_1$ the full electromagnetic field will be present.

These results have a consequence on the magnitude of the scattering
cross ection of light by cosmic strings. In Ref.~\cite{Witten} the cross
section per unit length of string at a frequency $\omega$ is
\begin{equation}
\frac{d\sigma}{dz}=\frac{\pi}{2\omega\ln (\Lambda /\omega)}\, .
\end{equation}
This will be modified for the portion of the electromagnetic field that
is screened.
\begin{equation}
\frac{d\sigma}{dz}=\frac{\pi}{2\omega}\left [\frac{\cos^2 2\theta_W}{\ln
(\Lambda /\omega)}+\frac{1-\cos^2 2\theta_W}{\ln (1/r_1\omega)}\right ]\, .
\end{equation}
{}\\

M.\ B.\ wishes to thank Professor D. Amati and Professor A. Schwimmer
for discussions and suggestions and the hospitality of SISSA, Trieste,
Italy, where part of this work was done. M.\ B.\ was supported in part by
the National Science Foundation under Grant PHY-9208386. H.\ R. was
supported by a SCIENCE EEC Astroparticles contract.

\end{document}